\documentstyle[12pt]{article}

\begin{document}

\begin{titlepage}

\pagestyle{empty}

\begin{flushright}
{\footnotesize Brown-HET-1009

January 1996}
\end{flushright}

\vskip 1.0cm

\begin{center}
{\Large \bf Inflationary Models Driven by Adiabatic Matter Creation}

\vskip 1cm

\renewcommand{\thefootnote}{\alph{footnote}}

L. R. W. Abramo$^{1,}$\footnote{e-mail:abramo@het.brown.edu},
J. A. S. Lima$^{1,2,}$\footnote{e-mail:limajas@het.brown.edu}  

\end{center}

\vskip 0.5cm

\begin{quote}
{\small $^1$ Physics Department, Brown University, 
Providence, RI 02912,USA.

$^2$ Departamento de F\'{\i}sica Te\'orica e Experimental,
     Universidade \\ 
$^{ }$ $^{ }$ Federal do Rio Grande do Norte, 
     59072 - 970, Natal, RN, Brazil.}
\end{quote}

\vskip 1.5cm

\begin{abstract}

\noindent The flat inflationary dust universe with matter creation 
proposed by 
Prigogine and coworkers is generalized and its
dynamical properties are reexamined. It is shown that the starting 
point of these models depends critically on a dimensionless parameter 
$\Sigma$, closely related to the matter creation rate $\psi$. For $\Sigma$ bigger or
smaller than unity flat universes can emerge, respectively, either like a Big-Bang
FRW singularity or as a Minkowski space-time at $t=-\infty$. The case
$\Sigma=1$ corresponds to a de Sitter-type solution, a 
fixed point in
the phase diagram of the system, supported by the matter creation
process. The curvature effects have also been investigated. The 
inflating 
de Sitter is a universal attractor for all expanding solutions regardless 
of the initial conditions as well as of the curvature parameter.

\end{abstract}

\end{titlepage}

\pagebreak
\baselineskip 0.8cm

\section{Introduction}

\hspace{.3in} Much effort has been spent to understand the effects of
the matter creation process on the universe evolution[1-7].
In the framework of the Friedman-Robertson-Walker (FRW) geometries, a
self-consistent phenomenological description for matter creation has
recently been proposed by Prigogine and coworkers\cite{Prig,PGN}. 
The leitmotiv of this approach is that the matter 
creation process, at the expense of the gravitational field, can happen only
as an irreversible process constrained by the usual requirements of
nonequilibrium thermodynamics. The crucial ingredient, however, is the 
explicit use of a balance equation to the number density of created 
particles in addition to the Einstein field equations(EFE). When 
properly combined with the second law of thermodynamics, such an 
equation leads 
naturally to a reinterpretation of the stress tensor, corresponding to 
an additional pressure term, which in turn, depends on the matter 
creation rate.

An extended manifestly covariant version of such a formulation has also
appeared in the literature\cite{LCW,CLW}. The nonequivalence
between the matter creation process and the mechanism of bulk viscosity, 
which has been widely used in
the literature as a phenomenological description to the former, has been
recently clarified\cite{LG,Gabriel}. In connection to this, we remark that
the general thermodynamic properties of the formulation by Prigogine and 
coworkers has
been more carefully investigated than its dynamic counterpart, which was
also presented in the mentioned papers as a new example of nontraditional
cosmology.

In this article we focus our attention on the cosmological
scenario proposed in Refs.\cite{Prig,PGN}. As
will be seen, our study will provide some general results often useful
to bring forward certain subtleties in the cosmological solutions which
apparently were not perceived at first by the mentioned authors.
To be more specific, in the scenario proposed in the quoted papers, the
spacetime starts from a Minkowski phase at $t=0$, with a particle number
density $n_o$ describing the initial fluctuation (Cf., for instance, 
\cite{Prig}, pg. 773). As will be shown here, however, there is 
no initial fluctuation since
the Minkowski spacetime starts at $t=-\infty$ with a number of
particles precisely equal to zero, as it should be. Subsequently, these 
models evolve smoothly  to a de Sitter phase,
in such a way that at $t=0$ all solutions describe an
expanding (inflationary) flat FRW-type universe. Another unnoticed
aspect is related to the existence of a new large class of solutions
starting from a big-bang FRW singularity and also approaching de Sitter
spacetime for late cosmological times. For all values of the parameter 
of the equation of state $p=(\gamma-1)\rho$ , the starting point of the 
flat universes depend critically on a 
dimensionless parameter closely related to the matter creation rate.
It can also be
shown that all expanding solutions converge to a 
flat inflating de Sitter solution. In other words, the latter 
is an attractor independent of initial 
conditions. 

Using phase space portrait techniques, the above analysis is 
extended to curved spacetimes. Due to degeneracy in flat space, the 
class of models proposed by 
Prigogine {\it et al.} is shown to split into two  distinct
classes. In particular, it is shown that only de Sitter and a space-like 
singularity (Big Crunch) are attracting nodes (stable points) in the phase 
space of solutions. For the first class, the former attractor is 
independent of the initial conditions, that is, all expanding solutions 
end up like a inflating de Sitter spacetime regardless of the values of
the curvature parameter. 

This paper is organized as follows: in Section 2, the basic equations
describing a FRW-type Cosmology with matter creation are presented.
In Section 3 we discuss the flat case for a matter content satisfying 
the $\gamma$-law equation of state. The flat dust model proposed by  
Prigogine {\it et al.} is discussed in detail and a second set of exact 
FRW-type solutions not foreseen by those authors are established. Finally, 
in Section
4 we develop a qualitative analysis in order to investigate the curvature
effects. The matter creation ansatz of Refs.\cite{Prig,PGN} is generalized and  
its dynamic consequences are discussed.

\section{Basic Equations}

Consider now the homogeneous and isotropic FRW line element
\begin{equation}
\label{line_elem}
  ds^2 = dt^2 - R^{2}(t) (\frac{dr^2}{1-k r^2} + r^2 d\theta^2+
      r^2sin^{2}(\theta) d \phi^2) \quad ,
\end{equation}
where $R$ is the scale factor and $k= 0, \pm 1$ is the curvature 
parameter. Throughout we use units such that $c=1$.

In that background, the basic dynamic equations for a self-gravitating 
perfect fluid endowed with matter creation reduce to \cite{CLW,LG}:

\vskip 0.2cm
\noindent {\it i}) Einstein's field equations(EFE):

\begin{equation}
\label{EE1}
    \chi \rho = 3 \frac{\dot{R}^2}{R^2}  + 3 \frac{k}{R^2} \quad ,
\end{equation}  

\begin{equation}
\label{EE2}
   \chi (p+p_{c})=-2\frac{\ddot{R}}{R}-\frac{\dot{R}^2}{R^2}
	             -\frac{k}{R^2} \quad ,
\end{equation}
whereas the energy conservation law,

\begin{equation}
\label{cons_en}
	\dot \rho + 3 ( \rho + p + p_c ) H = 0 \quad ,
\end{equation}
is the usual consistency condition contained in the independent 
EFE (\ref{EE1}) and (\ref{EE2}).
In the above equations, $\chi = 8 \pi G$ , $\rho$,
$p$ and $p_c=-\alpha \psi/3H$ are, respectively, the energy density, 
equilibrium and creation pressures. An overdot means time derivative and 
$H=\dot R / R$ is the Hubble parameter. 

\vskip 0.2cm
\noindent {\it ii}) The balance equation for the particle number
density:

\begin{equation}
\label{bal_n}
      \dot{n} + 3 n H = \psi \quad ,
\end{equation}
where $\psi$ is the particle source ($\psi >0$) or sink ($\psi < 0$).

The phenomenological parameter $\alpha$ appearing in the definition of
the creation pressure was
originally introduced in Refs. \cite{LCW,CLW}. As shown there, the Prigogine
et al. formulation is recovered in the ``adiabatic limit", namely when
$\psi$ is different from zero but the specific entropy (per particle) is
constant. From now on, we
consider just the ``adiabatic case'' for which the $\alpha$ parameter is given by (see Refs.\cite{CLW,LG})

\begin{equation}
\label{def_alpha}
\alpha=\frac{\rho + p}{n} \quad ,
\end{equation}
with the creation pressure assuming the following form

\begin{equation}
\label{pc_def}
p_c = - \frac{ \rho + p}{3 n H} \psi .
\end{equation}

In order to obtain a definite cosmological scenario with matter creation, one still needs to provide two additional
relations: the equation of state and the matter creation rate. In the
cosmological domain, the former is usually expressed in terms of the
``gamma-law" equation of state,

\begin{equation}
\label{g_law}
p=(\gamma -1)\rho \quad  \quad , \quad \quad 0 \leq \gamma \leq 2 \quad ,
\end{equation}
where $\gamma$ is the ``adiabatic index".

Note that, from (\ref{EE1}), (\ref{EE2}), (\ref{pc_def}) and (\ref{g_law}), 
the evolution equation for the scale factor $R$ can be written as

\begin{equation}
\label{Rdot}
     R\ddot{R} + (\frac{3 \gamma - 2}{2} - \frac{\gamma
     \psi}{2nH}) \dot{R}^2 + ( \frac{3 \gamma - 2}{2} -
     \frac{\gamma \psi}{2nH}) k = 0 \quad,  
\end{equation} 
which depends only on the matter creation rate $\psi$ and the number 
density of particles $n$.

Furthermore, using (\ref{cons_en}), (\ref{bal_n}) and 
(\ref{pc_def}), it is easy to
show that $n$ and $\rho$ satisfy the general
relation (see Ref.\cite{Nois})

\begin{equation} 
\label{n_rho}
n = n_{o}{ (\frac{\rho}{\rho_{o}}) }^{\frac{1}{\gamma}} \quad,
\end{equation}
so that, using the EFE (\ref{EE1}) one can obtain $n(R,H^2 )$, with
Eq.(\ref{Rdot}) now depending only on $\psi$. 

In what follows, we will first discuss the simple 
and exactly 
integrable flat case ($k=0$). This 
class includes Prigogine {\it et al.}'s dust model \cite{Prig,PGN}. The contributions of 
curvature terms will be analyze in section 4, but for now 
it suffices to say that flat solutions are not necessarily stable with
respect to curvature perturbations. In the language of dynamical systems,
FRW cosmology without matter creation is ``structurally stable" with
respect to curvature perturbations, while models with matter creation  are not.

\section{Flat Case: Two Classes of Models}

Let us now discuss the flat case. In terms of the Hubble 
parameter the evolution equation (\ref{Rdot}) for the scale factor reduces to: 

\begin{equation}
\label{Hdot_flat}
\dot H + \frac{3\gamma}{2} H^2 = \frac{ \gamma \psi}{2 n} H \quad ,
\end{equation}
while from (\ref{EE1}) and (\ref{n_rho}) the particle number density $n$ 
assume the form below,

\begin{equation}
\label{n_H_flat}
n = n_{o}(\frac{H}{H_{o}})^{\frac{2}{\gamma}} \quad .
\end{equation}

Now, in order to properly generalize the dust universe of 
Refs.\cite{Prig,PGN} 
we consider a class of models endowed with
the same matter creation rate

\begin{equation}
\label{rate_H2}
 \psi= 3 \beta H^2 \quad .
\end{equation} 
In the above expression $\beta$ is a new phenomenological parameter
($\frac{\alpha}{3}$ in the notation of Refs.\cite{Prig,PGN}). Parenthetically, since we are dealing 
with a spatially  flat spacetime, the above equation 
implies that $\psi$ is proportional to the energy density, more precisely, 
$\psi = \chi \beta \rho$. Of course, these 
relations 
are not equivalent in curved spacetimes, and so doing we are effectively 
working with two different creation rates (see section 4). 

Inserting Eqs.(\ref{n_H_flat}) and (\ref{rate_H2}) into (\ref{Hdot_flat}) one
obtains
\begin{equation}
\label{Hdot_flat_gd2}
\dot H + \frac{3 \gamma}{2} H^2 \left[ 1-(\frac{H}{H_d})^
	{\frac{\gamma -2}{\gamma}} \right] = 0 \quad \quad 
	( \gamma \neq 2 ) \quad ,
\end{equation}
and

\begin{equation}
\label{Hdot_flat_g=2}
\dot H + 3 (1-\frac{\beta H_o}{n_o}) H^2 = 0 \quad \quad
	(\gamma = 2) \quad .
\end{equation}

In Eqs.(\ref{n_H_flat})-(\ref{Hdot_flat_g=2}), $n_o$ is the value of
$n$ for $H=H_o$ and $H_d$ is given by

\begin{equation}
\label{def_Hd}
H_d = H_o { ( \frac{n_o}{\beta H_o} ) }^{\frac{\gamma}{\gamma-2}} \quad 
	\quad (\gamma \neq 2) \quad .
\end{equation}
The case $\gamma=2$ is the simplest one to be analyzed. However, since
it does not exhibit the peculiarities present in the class of solutions
of (\ref{Hdot_flat_gd2}), which contains the model of Refs. 
\cite{Prig,PGN}, we do not consider it in what follows.
Apart from some interpretation 
problems, Prigogine's et al. model must
be recovered by taking $\gamma=1$ in the general solution of (\ref{Hdot_flat_gd2}). A straightforward integration of this later equation yields

\begin{equation}
\label{H_sol}
H=H_d ( 1 + A R^{\frac{3(\gamma-2)}{2}} )^{\frac{\gamma}{2-\gamma}} \quad ,
\end{equation}
where $A$ is a dimensional integration constant.

Now, in order to introduce in our approach a useful dimensionless parameter,
we compute the value of $A$ as a function of $H_d$ , $H_o$ and $R_o$ , where 
$R_o$ is the value of $R$ when $H=H_o$. From (\ref{H_sol}) one reads

\begin{equation}
\label{def_A}
A= ( \Sigma -1 ) R_o^{\frac{3}{2} (2-\gamma) } \quad ,
\end{equation} 
where 

\begin{equation}
\label{def_Si}
  \Sigma=\frac{n_o}{\beta H_o} \quad .
\end{equation}
In terms of $\Sigma$ the Hubble parameter takes the form

\begin{equation}
\label{H_Si}
H = H_d \left[ 1 + (\Sigma-1)(\frac{R}{R_o})^{\frac{3}{2}(\gamma-2)} \right]^
	{\frac{\gamma}{2-\gamma}} \quad ,
\end{equation}
and from (\ref{n_H_flat}) and (\ref{H_sol})-(\ref{def_Si}), the particle
number density is given by

\begin{equation}
\label{n_Si}
n=n_o {\Sigma}^{\frac{2}{\gamma-2}}
      \left[ 1 + (\Sigma-1)(\frac{R}{R_o})^{\frac{3}{2}(\gamma-2)} \right]
      ^{\frac{2}{2-\gamma}} \quad ,
\end{equation}
with the time-dependent net number of particles, $N=nR^{3}$, assuming the following form: 

\begin{equation}
\label{Number}
N(t) = n_o R^3 \Sigma^{\frac{2}{\gamma-2}} \left[ 1-
	(\Sigma-1)(\frac{R}{R_o})^{\frac{3}{2}(\gamma-2)} \right]
	^{\frac{2}{2-\gamma}} \quad .
\end{equation}

For these models, the qualitative behavior at early and late times  can be 
easily determined 
from Eqs.(\ref{H_Si}) and (\ref{n_Si}) or (\ref{Number}). If $\Sigma=1$, the Hubble 
parameter
is constant, $H=H_d$ , with the solutions 
reducing to a flat de Sitter-type
universe regardless of the value of 
$\gamma$. In this way, we may have, for
instance, a 
de Sitter universe supported by the adiabatic creation of photons
($\gamma=4/3$) or dust ($\gamma=1$). Such a fact had 
already been observed 
in Ref.\cite{Prig}. Indeed, $H=H_d$
is a special solution, a singular point in the phase diagram of the system
$\{\dot H (\rho ,H)$, $\dot \rho (\rho ,H)\}$
(see Eqs.(\ref{Hdot}) and (\ref{rhodot}) in section 4). It should be noticed that 
for large
values of $R$, that is, $(R/R_o)^{-3(2-\gamma)/2} \ll 1$, the models evolve
to this inflating solution. 

How these flat universes emerge depends only on the sign of $A$ or
equivalently, if the dimensionless 
parameter $\Sigma$ is bigger, smaller or
equal unity. For $\Sigma>1$, the second term on the rhs of (\ref{H_Si}) is 
dominant for
small values of $R$, that is, $(R/R_o)^{-\frac{2-\gamma}{\gamma}} \gg 1$,
thereby leading to the usual FRW singularity, $R \sim t^{2/3\gamma}$ .
Since $H \rightarrow \infty$ in this limit, it follows from 
(\ref{n_rho}) and (\ref{n_H_flat}) that $\rho$ and $n$ are infinite as
$t$ goes to zero. These results characterize a class of singular
solutions not previously 
perceived in Refs.\cite{Prig,PGN}.

On the other hand, if $\Sigma < 1$ it follows from (\ref{H_Si}) that there
is a minimal value of $R$, namely: 

\begin{equation}
\label{R_min}
R_{min} = R_o (1-\Sigma)^{\frac{2}{3(2-\gamma)}} \quad ,
\end{equation}
for which $H=0$. In addition, Eqs.(\ref{n_rho}), (\ref{n_H_flat}) and
(\ref{n_Si}),  
yield
$\rho = n = N = 0$ at $R=R_{min}$, making explicit that such models start, 
for all values of $\gamma$, as a Minkowski vacuum. 
We remark that, from
(\ref{pc_def}) and (\ref{n_H_flat}),
the creation pressure $p_c$ scales with $H^{3\gamma-2}$ . Thus, 
according to the above results, in the
beginning of the universe the creation pressure was either zero or
infinite if, respectively, $\Sigma >1$ or $\Sigma<1$. Only in the de Sitter
case ($\Sigma=1$) the creation pressure has a finite(and constant) value.

The above results are rather important in what follows. For $R=R_o$
one obtains $H=H_o$, $n=n_o$ and $N=n_{o}{R_{o}}^{3}$ for all values of 
$\gamma$ and $\Sigma \neq 1$.
For $\Sigma=1$, (\ref{H_Si}) and (\ref{n_Si}) reduce to the constant values 
$H=H_d$
and $n=n_o$ as required by the symmetries of the de Sitter spacetime. 
The latter case is different from the de Sitter phase attained in the 
course of the evolution for $\Sigma \neq 1$. In fact, for $R \gg R_o$ , 
(\ref{n_Si}) yields

\begin{equation}
\label{n_lim}
n_d = n_o \Sigma^{-\frac{2}{2-\gamma}} \quad,
\end{equation}
which may also be obtained from (\ref{n_H_flat}) taking $H=H_d$. The important
point to keep in mind here is that models starting as Minkowski ($\Sigma < 1$) have $n=0$ when $R=R_{min}$, that is, $n_o$ cannot be the number density
at the beginning of the universe. 

Therefore, since for $\gamma=1$ the
energy density is $\rho=n M$ (where $M$ is the mass of the created dust particles)
we can say that the authors of Refs.\cite{Prig,PGN} studied
only the case $\Sigma < 1$ , that is, 
$M > \frac{3 n_o}{\chi \beta^2}$. Note that $n_o$ and $\beta$ are coupled to 
give the
natural mass scale of the models so that for $\gamma=1$ the condition
$\Sigma \leq 1$ ($\Sigma \geq 1$) assumes the interesting form $M \geq 
M_c $ ($M \leq M_c$), where the critical
mass is $M_c = \frac{3 n_d}{\chi \beta^2}$, just the mass of the particles created in the de Sitter spacetime.  For this steady state scenario, we see
from  (5) and (13) that $\beta=n_d{H_d}^{-1}$, and using the present
data it is readily obtained  $M_c \approx 1Gev$ i.e., the proton mass. Naturally, the above condition may also be translated as a constraint on $\beta$, the free parameter of the models. In terms of the mass $M$,
the critical value of this parameter is given by $\beta_c= \frac{3H_d}{ 8\Pi}(\frac {M_{pl}^{2}}{M})$, where $M_{pl}$
is the Planck mass.   
In contrast with the approach developed here, the mass  of the 
created particles in Refs.\cite{Prig,PGN} has been estimated 
by assuming that the de Sitter phase is 
unstable and evolves continuously (up to first derivatives) to the FRW 
radiation phase. 

\subsection{ Prigogine {\it et al.}'s Model ($\Sigma < 1$, $\gamma = 1$) }

Let us now analyze with more detail the case $\gamma=1$. As remarked above, 
in this case $\chi \rho_o = \chi n_o M = 3 H_o^2$ , so that 
(\ref{def_Hd}) and (\ref{n_lim}) reduce to

\begin{equation}
\label{Hd_3}
H_d = \frac{\chi \beta M}{3} \quad ,
\end{equation}
and

\begin{equation}
\label{n_3}
n_d = \frac{\chi \beta^2 M}{3} \quad .
\end{equation}

The aforementioned results are presented,
respectively, in Eqs.(20) and (4) of Refs.\cite{Prig,PGN}. 
Now, in 
order to relate (\ref{Hd_3}) and (\ref{n_3}) with the characteristic time 
scale of the de Sitter phase we insert (\ref{Hd_3}) into (\ref{H_Si}) and 
integrate it with $\gamma = 1$ and $\Sigma < 1$ to obtain

\begin{equation}
\label{R_3}
R(t) = R_o {[ 1 + {\Sigma}
	( e^{\frac{\chi \beta M}{2} t} -1 ) ]}^{2/3} \quad ,
\end{equation}
which is the same expression for the scale function presented in Refs.\cite{Prig,PGN}.
Therefore, after a characteristic time $\tau_c = 
\frac{2}{\chi \beta M} = \frac{2}{3} H_d^{-1}$, the universe reaches the
de Sitter phase expanding as

\begin{equation}
\label{R_3_lim}
R(t) \simeq R_o (\frac{n_o}{n_d})^{1/2} e^{H_d t} 
\quad ,
\end{equation}
where $n_d$ and $H_d$ were defined by (\ref{Hd_3}) and (\ref{n_3}). On the
other hand, it follows from (\ref{R_min}) that the minimal value of
$R$ for $\gamma=1$ is given by

\begin{equation}
\label{R_min_3}
R_{min} = R_o [ 1- \Sigma ]^{2/3} \quad ,
\end{equation}
which depends on $n_o$ through $\Sigma$. Hence, if we consider $R_o=1$ 
as in Refs.\cite{Prig,PGN}, one has $R_{min} < 1$, with 
(\ref{R_min_3}) making explicit the constraint $\Sigma < 1$ as noticed 
earlier. Of course, the natural choice is to take not $R_o$ but 
$R_{min}=1$. However, the important step 
is to determine at what time the scale 
factor assumes its minimal value 
$R_{min}$. From (\ref{R_3}) we see that $R=R_{min}$ only if $t=-\infty$, 
which is
coherent with our qualitative analysis. Note that for $t=0$ (\ref{R_3}) 
yields
$R=R_o$ and, (\ref{n_Si}), $n=n_o$. In this context we remark that the authors
of Refs.\cite{Prig,PGN} truncated arbitrarily the time 
coordinate at $t=0$, and since the arbitrary value of $R_o$ has been 
fixed equal to
unity, they obtained $R(0)=1$ ;
in their words: ``the universe emerges without singularity at $t=0$, with a
particle number density $n_o$ describing the initial Minkowskian fluctuation"
(see Ref.\cite{Prig}, pg.773). Indeed, for $t=0$, in 
addition to
$R=R_o$ and 
$n=n_o$ , one obtains from (\ref{n_H_flat}) or equivalently 
from (\ref{def_Hd}), (\ref{def_Si}) and (\ref{H_Si}) that the Hubble parameter 
itself is $H=H_o$ , making it clear that the spacetime is not Minkowski.
It thus follows that at $t=0$ the solution (\ref{R_3}) describes an 
expanding
FRW-type universe driven by the matter creation process. Indeed,
such a result is valid for all values of $\gamma$ in the considered interval 
and $\Sigma \neq 1$. As a matter of fact, the extension of the time coordinate for $t= -\infty$ may also be justified looking for the expression of the
total number of particles given by (\ref{Number}).  
By taking $\gamma=1$ in the later equation and replacing the 
expression of $R$ given in (\ref{R_3}) one obtains

\begin{equation}
\label{Number_lim}
N(t)=N_o e^{\beta M t} \quad ,
\end{equation}
which is the same expression derived in Refs. 
\cite{Prig,PGN} using a different approach. Notice that the Minkowskian limit,
$N \rightarrow 0$, is recovered only if $t \rightarrow -\infty$ in accordance with our previous
comments, while for $t=0$ or equivalently $R=R_o$ one has $N=N_o$, as it should be.
Despite the fact that this ``initial fluctuation'' $n_o$ does not exist, it 
will be shown that there is a structural instability related with these 
types of models, which may be discussed using the dynamical systems technique
(see section 4).

Summarizing, this class of spacetimes ($\Sigma < 1$), which includes the 
dust case, emerges at $t=-\infty$ as
a true Minkowski spacetime, that is, $\rho=n=H=0$ and $R=R_{min}$, with 
a general expression 
given by (\ref{R_min}). There is no initial fluctuation. Due to the matter creation process, the 
universe evolves smoothly from
Minkowski to a de Sitter spacetime. This is an interesting example of 
spacetimes which are 
unbounded in time, that is, their evolution 
ranges the time interval (${-\infty, \infty}$) and 
are also free of 
physical singularities. Naturally, the 
characteristic quantities of the de Sitter 
phase may, for
certain special values of $\gamma$, be 
independent of $n_o$ ; a particular
value of $n$ arbitrarily fixed at $t=0$. This 
is not remarkable and happens
just for $\gamma=1$ as a consequence of the relation $\rho=nM$ (see 
Eqs.(\ref{Hd_3}), (\ref{n_3}) and the definition of $\tau_c$). 

\subsection{A New Model ($\Sigma > 1$, $\gamma=1$)}

Having in mind that the case $\Sigma=1$ is trivial we now analyze the last 
possibility, that is, $\Sigma >1$ or $\chi \beta^2 M < 3 n_o$ . In this 
case, as
expected, Eq.(\ref{H_Si}) is not modified, but on 
the other hand there is 
no $R_{min}$ , so that $R \rightarrow 0$ and 
$H \rightarrow \infty$, characterizing an initial FRW-type singularity.
Taking into account such remarks in the integration process of (\ref{H_Si}),
it is easily found, for $\gamma=1$,

\begin{equation}
\label{R_4}
R(t) = R_o [ (\Sigma -1) (e^{\frac{3}{2} H_d t} -1) ]^{2/3} \quad .
\end{equation}

This expression is consistent with our earlier qualitative analysis. For times
$t > \tau_c = \frac{2}{3} H_d^{-1}$ , the model approaches the de
Sitter regime. If $t \ll \tau_c$ then (\ref{R_4}) yields

\begin{equation}
\label{R_4_2}
R(t) \simeq R_o [ (\Sigma -1) ( 1+ \frac{3}{2} H_d t + \cdots -1) ] ^{2/3}
\quad ,
\end{equation}
so that

\begin{equation}
\label{R_4_3}
R(t) \simeq R_o [ \frac{3}{2} H_d (\Sigma -1) t ] ^{2/3} 
\quad .
\end{equation}

This was expected, since for large values of $\Sigma = H_o/H_d$, 
the equation above approaches 
$R \sim  R_o (\frac{2}{3} H_o t)^{2/3}$, which is the standard form of
the FRW dust model\cite{Weinberg}.

\section{General Case: Qualitative Analysis}

The flat case have already showed striking richness,
and a natural extension to curved spacetimes is
already more than justified. However, since is rather difficult to 
analytically derive the complete set of solutions, we will employ a 
different approach provided by the qualitative analysis of dynamical 
systems \cite{Andronov,Bogoyav}. The basic idea 
is to reduce the 
field equations to a bidimensional autonomous system and  perform the 
qualitative analysis of all solutions. As we know, the first step is set 
up a convenient set of dynamic variables, which we choose as being the 
energy density $\rho$ and the Hubble parameter $H$. Using
the definition $H= \dot R / R$ and the 
constraint (\ref{EE1}) we rewrite (\ref{Rdot}) as

\begin{equation}
\label{Hdot}
	\dot H =  \frac{\gamma }{2} (\frac{\psi}{3 n H} - 1) \chi \rho
		 + \frac{1}{3}(\chi \rho - 3H^2) 
\quad .
\end{equation}

The second dynamical equation is just the energy conservation 
law(\ref{cons_en}). Inserting (\ref{pc_def}) and (\ref{g_law}) it takes 
the following form

\begin{equation}
\label{rhodot}
	\dot \rho =  3 \gamma (\frac{\psi}{3 n H} - 1 ) \rho H \quad .
\end{equation}
Since the ratio ${\Psi}/{n}$ is a function 
of the energy density or 
Hubble parameter (or some combination of them) our dynamical system 
is fully defined by Eqs.(\ref{Hdot}) and  (\ref{rhodot}). 
In what follows we consider only a positive-definite $\gamma$.
First of all, 
we remark that the flat solutions corresponding to 
$\chi \rho = 3 H^2$ are reconstructed in the phase space irrespectively 
of the matter creation rate. In fact,  
taking the ratio of (\ref{Hdot}) to (\ref{rhodot}) we see
that the parabola $\chi \rho = 3 H^2$ corresponds to 
the solutions for the flat case. Therefore, such    
a curve is a separatrix for the general 
solutions even in presence of matter creation. This 
is just a reminder that (\ref{EE1}) constrains open 
and closed spacetimes in such a way that they cannot evolve 
into one another. In general, there are at least two
singular points ${\dot H= \dot \rho= 0}$ in the phase plane: 
 
\begin{equation}
\label{Mp}
[H,\rho]=[0,0] \quad,  \end{equation}  
\begin{equation}
\label{MP1}
[H, \rho]=[H_d, \rho_d] \quad,
\end{equation} 
which correspond to Minkowski and de Sitter spacetimes. As will be 
shown ahead, the former plays a completely different role in this new 
context, while the later does not exist in the 
absence of matter creation.
 
A phase diagram (or portrait) is a plot of the solutions to the 
dynamical system in the plane of the variables, with a flow corresponding 
to the arrow of time.
As we know, this kind of graph can be 
compactified by changing the variables in order to  
bring infinity to a boundary in the phase portrait of the 
transformed dynamical system. A
particularly useful new set of variables may be 
defined as the conformal mapping\cite{Oliveira}

\begin{equation}
\label{Ht_gen}
	H =  H_d \frac{r}{1-r} \sin{\phi}
\end{equation}
and

\begin{equation}
\label{rhot_gen}
        \chi \rho = 3 H_d^2 { \left( \frac{r}{1-r} cos{\phi} \right) }^2 
	\quad ,
\end{equation}
where $H_d$ is an arbitrary scale that will be set equal to the 
previously defined Hubble parameter of the de Sitter spacetime.
The inverse transformation is easily computed as

\begin{equation}
\label{r_gen}
	r = { \left[ 1 + 
\frac{H_d}{(\frac{\chi \rho}{3} + H^2)^{\frac{1}{2}}} \right] }^{-1}  
	\quad , \end{equation}

\begin{equation}
\label{phi_gen}
	\phi = \arctan{ \frac{3 H^2}{\chi \rho} } \quad .
\end{equation}

It is readily seen that infinite values of $\rho$ and $H$ are brought onto the 
circle of radius $r=1$. In terms of the new variables $\{r,\phi\}$, we can
recast the 
dynamical system (\ref{Hdot})-(\ref{rhodot}) as

\begin{equation}
\label{rdot}
\dot r = r H_d { \left\{ 
	\sin{\phi} \cos{2 \phi} - 
	\frac{3 \gamma}{2} \cos{\phi} \sin{2 \phi} {\left[ 
	1- \frac{(1-r)}{r} \frac{ \psi }{ 3 n H_d \sin{\phi} } 
	\right] }  \right\} }
\end{equation}
and 

\begin{equation}
\label{phidot} 
\dot \phi = \frac{r}{1-r} H_d 
        \cos{\phi} \cos{2 \phi} { \left
\{ 1- \frac{3 \gamma}{2} {\left[
        1-\frac{(1-r)}{r} \frac{ \psi }{ 3 n H_d \sin{\phi} }  
	\right] } \right\} } \quad .
\end{equation}

The phase portrait of usual FRW models ($\psi =0$; see, e.g., 
Ref. \cite{Belinski}) 
is conformally transformed into the diagram of Fig.1, which have 
been shown for further 
comparison. The parabola $\chi \rho = 3 H^2$ 
in the original phase 
space is mapped onto the radiuses at $\phi = \pm \pi/4$, and 
the Big Bang (Big Crunch) singularities, respectively 
$(\rho, H) = (\infty, \pm \infty)$, 
are tamed to the points $r=1$, $\phi=\pm \pi/4$.  
The shape of solutions is kept intact near the origin of the transformed graph. As 
expected, for these simple
FRW models we are able to find exact expressions to the integral curves 
shown in Fig.1, since we
know these systems to be integrable (see, e.g., Ref. \cite{AssadL}). By 
taking the
ratio of (\ref{rdot}) to (\ref{phidot}) with $\psi=0$ we 
immediately solve the 
equation $dr/d\phi$ to find the solution

\begin{equation}
\label{sol_FRW}
r(\phi) = 1 - (1-r_o) {\left[
	|\cos{\phi}| {| \cos{2 \phi} |}^{-\frac{3 \gamma}{4}} 
	\right]}^{\frac{2}{2-3\gamma}} \quad ,
\end{equation}
where $r_o$ and $r$ are constrained to the interval $[0,1]$. Note that this 
formula is valid even for the unphysical region
$\rho < 0$ ($\cos{\phi} < 0$) which is also depicted for
completeness. This will later prove helpful in order to analyze the nature
of equilibrium points such as the Minkowski spacetime at the origin $r=0$ 
($\rho=H=0$), and the physical singularities at $r=1$ and $\phi=\pm\pi/4$ 
(Big Bang and Big Crunch respectively).

At this point we should recall that although in flat spacetimes the energy
density and Hubble parameter are indistinguishable, in curved 
spacetimes this is not so. In this way, the ansatz proposed in 
Refs.\cite{Prig,PGN}, 
$\psi_{\rm flat}=3\beta H^2$, must be somewhat generalized to 
include the presence of curvature. In principle, if we mimic  
the bulk viscosity mechanism, a more reasonable phenomenological law for 
$\Psi$ should be a power-law dependence $\Psi=\eta \rho^{\nu}$, where $\eta$ 
is a dimensional constant and $0 \leq \nu \leq 1$ (see, for instance, 
Ref.\cite{Belinski}).  In this case, 
regardless of the curvature parameter, the 
Prigogine et al. ansatz corresponds to $\nu=1$
and $\eta=\beta \chi$. However, this kind of phenomenological law does not include the case
$\Psi= 3\beta n H$ ($\beta$ constant), discussed in Ref.\cite{Nois}. 
In this connection, we recall that the matter creation rate is a degree 
of freedom introduced in 
the theory trough the balance equation for the particle number density. In this 
way, we assume here a rather general expression which include both 
cases, namely:

\begin{equation}
\label{psi_gen}
	\psi = 3 \bar \beta n {\left( \frac{\chi\rho}{3} \right) }^{\mu/2}
		H^{\nu + 1} \quad .
\end{equation}
where $\bar \beta$ is a constant.

As noted above, a general feature of models with matter creation 
is the presence of a de Sitter solution.
It is easy to see how these solutions come up: 
consider the curve $\cal C (\rho ,\rm H)$ corresponding to the
equality $\psi (\rho ,H)=3nH$ which, from 
Eqs.(\ref{Hdot})-(\ref{rhodot}) 
always crosses the parabola $\chi\rho=3H^2$ (or, equivalently, the straight 
line $\phi=\pi/4$ of Fig.1). At that precise 
point $[H_d,\rho_d]$ 
(hereafter indicated as $\cal I$), it is readily seen that both $\dot \rho$ 
and $\dot H$ 
are zero, meaning an exponentially expanding universe with constant  
density, i.e., an 
inflating universe. The question of whether models evolve towards it or
outwards from it is related to the nature of the singular point 
$\cal I$. This is done by finding the Liapunov coefficients of the
linearized dynamical system at that point \cite{Andronov}. For the matter 
creation rate of the form assumed above,
after some simple algebra we find 

$$
\lambda_1 =-2
$$
and 

$$
\lambda_2 = \frac{3 \gamma}{2} ( \mu + \nu ) \quad .
$$
This means that the fixed point $\cal I$ is either a saddle 
point, an attracting Jordan node or a simple attracting node (attractor), if
$\lambda_2 > 0$, $\lambda_2 = 0$ or $\lambda_2 < 0$ respectively. The 
absence of complex Liapunov coefficients for any $\gamma$ or $\beta$ 
guarantees no closed orbits around
$\cal I$, which is obvious given the 
constraint (\ref{EE1}) and the 
discussion after Eq.(\ref{rhodot}). 
The reader may verify for himself 
that $\lambda_2$ is negative if we consider 
$\Psi= \beta \rho$, since in this case $\nu=-1$ and 
$\mu=2(\gamma-1)/\gamma \leq 1$. It thus follows that the inflating 
universe is an attractor for the class of models considered
in Refs.\cite{Prig,PGN}. The phase portrait is 
shown in Fig.2. As explained, the point 
$\cal I$ is an attractor, and all expanding 
solutions converge there. 
Since $H=0$ ($\cal H$ in Fig.2) is a line 
of singular points, expanding 
and contracting models are never in contact. 
Such an effect is probably due to 
the creation pressure
$p_c$ which diverges at that 
line (see Eq.(\ref{pc_def})). Actually, it is easy 
to see that all along line 
$\cal H$ the rate $\dot H$ (see Eq.(\ref{Hdot})) 
goes to infinity.

In this model, the expanding universe seems 
to have appeared either 
from a FRW-type singularity at $\cal B$ ($r=1$, $\phi=\pi/4$), or at
any point in the singular line $\cal H$. In other words, all expanding 
spacetimes evolved from a singularity of some kind. Even Minkowski 
(denoted $\cal M$), which is a nonsingular saddle point in FRW 
models (see fig.1), is a 
multiple equilibrium singular point here. Note also that closed 
universes show no turning points, either 
expanding or contracting forever.

Contracting universes in this scheme can begin either from Minkowski or 
from the same singular line as the expanding ones, $\cal H$. All 
solutions now evolve to a crunch, exactly as in the FRW case.
In the region of negative energy (left half of Fig.2), contracting 
universes emerge 
either from the singular line $\cal H$ or from $\cal M$, and finish
with infinite negative energy at ($r=1$, $\phi=-\pi$).
Expanding universes with negative energy all start from $
\cal H$, and 
end up as flat empty universes at $\cal M$. In sum, if 
the universe described by this model started from a fluctuation with both
$H>0$ and $\rho > 0$ , 
it would evolve towards the inflating universe at $\cal I$.
On the other hand, if it started from a curvature fluctuation 
($H \neq 0$, $\rho = 0$), it would certainly end up either in a crunch or 
in an empty Minkowski spacetime. 
The case of a matter fluctuation ($H=0$, $\rho \neq 0$) is somewhat
ill-defined, since $p_c$ diverges on the line $\cal H$.

\section{Conclusion}

In this paper, by allowing both pressure and curvature, we have generalized 
the flat pre-inflationary stage with 
matter creation proposed by Prigogine and coworkers. 
All expanding solutions evolve to the flat de Sitter spacetime in
the infinite cosmic time of their evolution.
Physically, open and
closed models are always
singular, while the flat solutions split into two 
distinct subclasses which depend on the matter creation rate.  The beginning of 
the universe in this case may be either a Minkowskian vacuum 
or a FRW type 
singularity. The avoidance of a singularity depends on 
the strength of the matter creation process, which has been 
translated here as a condition on the dimensionless $\Sigma$ parameter.
It thus appear that a certain degree of fine tuning is needed in order
to escape from a physical singularity in these models.

\section{Acknowledgments}

It is a pleasure to thank R. Brandenberger, A.
Sornborger and N. Lemos for valuable suggestions and a critical reading
of the manuscript. JASL is particularly grateful for A. Germano for many 
discussions in the beginning of this work. 
One of us (JASL) is grateful for 
the hospitality of the Physics
Department of Brown University. This work was partially 
supported by the Conselho Nacional de 
Desenvolvimento Cient\'{\i}fico e
Tecnol\'ogico - CNPq 
(Brazilian Research Agency), and by the US 
Department Of Energy under grant 
DE-FG02-91ER40688, Task A.

\newpage

\begin{center}
{\bf Captions for Figures
\vskip 1cm
Fig.1}
\vskip 0.5cm
\end{center}
\it
  
\begin{quote}
{\small {\bf Fig. 1} - Phase diagram of the  standard FRW models mapped onto a 
disk. Flat universes are represented by the straight lines 
at $\phi= \pm \pi/4$. The origin $\cal M$ is an 
attracting node. Point $\cal B$
corresponds to the Big Bang singularity, while $\cal C$ is the Big Crunch
at the end of the evolution of closed universes. } 
\end{quote}

\vskip 1.0cm

\begin{center}
{\bf Fig. 2}
\end{center}

\vskip 0.5cm

\begin{quote}
{\small {\bf Fig. 2} - Phase diagram of a model with $\psi = \beta \chi \rho$. 
Again flat universes corresponds to straight lines at $\phi= \pm \pi/4$. However, the point $\cal I$ is an attractor for all expanding solutions. 
$\cal H$ (corresponding to $H=0$, or $\phi=0$) is a line 
of singular points and 
Minkowski ($\cal M$) now is a Saddle point.} 
\end{quote}

\end{document}